\documentclass{aa}
\usepackage{graphicx}
\usepackage{amsmath}

\begin{document}

    \title{Number Counts of Bright Extremely Red Objects: Evolved
    Massive Galaxies at $z\sim1$}  

   \author{P. V\"ais\"anen \inst{1,2} \and P.H. Johansson \inst{3,4}}

   \offprints{P. V\"ais\"anen\\
   \email{pvaisane@eso.org}}

   \institute{European Southern Observatory, Casilla 19001, Santiago,
   Chile  \and Departamento de Astronom\'ia, Universidad de
   Chile, Casilla 36-D, Santiago, Chile \and Institute of Astronomy,
   Madingley Road, Cambridge, CB3 0HA, UK \and Observatory, P. O. Box 14, FIN-00014 University of
   Helsinki, Finland
}
   
\date{Received / Accepted }
\authorrunning{V\"ais\"anen \& Johansson}
\titlerunning{Bright EROs and massive galaxies at $z\sim1$}

\abstract{We present results on number counts of Extremely Red Objects
  (EROs) in a 2850 arcmin$^2$ near-infrared survey performed in
  European Large Area ISO Survey (ELAIS) fields at $K<17.5$.  Counts
  of EROs are extended to brighter levels than available previously,
  giving $0.002 \pm 0.001 \rm arcmin^{-2}$ at $K<16.5$ and consistent
  numbers with literature values at fainter magnitudes.
 Photometric redshifts from HYPERZ as well as GRASIL
  model SEDs of galaxies imply that our EROs are located in the range
  $z=0.7-1.5$, with the bulk of the population at $z\sim1$. 
  Taking advantage of the ISO data in the fields, we use mid-IR
  detections to constrain the number of 
  dusty EROs, and also discuss the superior capabilities of Spitzer
  Space Telescope to
  detect dusty EROs. Both the mid-IR data and the use of colour-colour
  diagrammes indicate that at most 10-20\% of the EROs in this bright
  regime are dusty starbursting systems.  The space density of
  our EROs, interpreted to be counterparts of local $>2-3L^{\star}$
  massive galaxies at around $z\sim1$, is estimated to be
  $\approx2\times10^{-5} \ {\rm  Mpc^{-3}}$, which is consistent with
  local values.  Furthermore, the cumulative number counts at our
  bright magnitudes are remarkably well fitted by pure luminosity
  evolution models. 

}

\maketitle

\section{Introduction}


Extremely Red Objects (EROs, selected for example by $R-K>5, I-K>4$
colours) 
have received much focus recently by virtue of their potential as a
powerful window into the galaxy formation era.  By and large,
the majority of EROs have been incorporated into a bimodal population,
where the extreme red colours are attributed  
either to old passively evolving distant $(z>1)$ elliptical galaxies
or to extremely dust reddened starburst galaxies (see e.g.\ recent
papers by 
Cimatti et al. \cite{Ci03}, Wold et al. \cite{Wo03}, Yan \& Thompson 
\cite{Ya03}, Takata et al. \cite{Ta03}, Daddi et al. \cite{Da02},
Smail et al. \cite{Sm02}, Roche et al. \cite{Ro02}, and 
references therein for earlier pioneering 
ERO surveys).  It is the class of
aged early type galaxies whose number densities 
provide the strongest constraints on models of galaxy evolution.
On the other hand, the dusty EROs could be related to the (ultra)
luminous IR-galaxies producing the bulk of the total energy in the
Universe since the recombination era (see e.g.\ Elbaz \& Cesarsky
\cite{El03} for a review). 


The dusty ERO population contaminates the otherwise, in principle,
clean samples of high redshift massive ellipticals. This is the
reason why the separation of elliptical EROs from dusty EROs, using
photometric, spectroscopic, and morphological analyses, has
received the main attention of ERO studies recently.  In general, a large
number of massive galaxies at redshifts of unity and over would support
a traditional monolithic collapse scenario where galaxies form at
high redshift 
in a single collapse and then evolve passively (e.g. Eggen et
al. \cite{Eg62}; Larson \cite{La75}).  Significantly smaller
numbers of massive high-$z$ ellipticals on the other hand fit
predictions from the hierarchical assembly scenarios, where massive
galaxies are formed only 
relatively recently from the merging of smaller units 
(e.g.\ White \& Rees \cite{Wh78}; White \& Frenk \cite{Wh91};
Somerville \& Primack \cite{So99}; Cole et al. \cite{Co00}). 

Since EROs are extremely faint optically $(R\ge 24)$, their redshifts
and spectral and morphological properties have remained largely
unknown until very recently.   
Cimatti et al. (\cite{Ci02}) carried out a VLT 
spectroscopic survey of 45 EROs with $R-K>5$ and $K_{s}<19.2$. 
They identified approximately $1/3$ of the EROs 
as old elliptical systems, $1/3$ as dusty starburst galaxies, while
$1/3$ \ remained  
unidentified. The mean redshift was found to be $z\sim 1.0$ for both
populations.  Yan, Thompson, \& Soifer (\cite{Ya04}) on the other hand
find a small 10-15\% fraction of isolated passive systems. 
Yan \& Thompson (\cite{Ya03}),
Cimatti et al. (\cite{Ci03}), Moustakas et al. (\cite{Mo04}), and
Gilbank et al. (\cite{Gi03}) have
used HST morphologies to differentiate 
between the classes: the results have further complicated the picture,
as a large fraction (25-65\%) of EROs seem to be disks (though see
also Moriondo et al.\ \cite{Mo00}).  The 
expectation originally had been a clearer distinction into spheroids
on one side, and irregular and interacting types on the other,
expected for extremely dusty star forming galaxies.  
The relative fractions of different types of EROs thus still
remain uncertain (due to for example different selection criteria
used) and the results of testing surface densities against galaxy
formation scenarios are inconclusive. 

In addition to spectroscopic and morphological methods mentioned
above, one may separate the old elliptical and dusty EROs 
by means of longer wavelength observations:
any ERO detected at mid-IR to radio wavelengths should belong to the
dusty population.  
Systematically this has been attempted by Mohan et al.
(\cite{Mo02}) in the sub-mm to radio and Smail et al.\ (\cite{Sm02})
in the radio.  
Similar fractions (albeit with wide spread) as referred to above of
dusty EROs were found in the latter study, while Mohan et
al. (\cite{Mo02}) find a much lower fraction.

In this paper we attempt to separate EROs, for the first time, 
based on their mid-IR properties.  As with far-IR to radio methods,
this is a clear-cut definition, since the difference of
ellipticals and starbursting galaxies is very large - distant evolved
ellipticals would not be detected 
with ISO whereas dusty EROs should have strong mid-IR flux.
Also, mid-IR
avoids the identification problems of e.g.\ large sub-mm beams.
We have surveyed an ISO/ELAIS (European Large 
Area \textit{ISO} Survey; Oliver et al. \cite{Ol00}) field in 
the near-IR and matched the results with an optical dataset.  We make
use of the newly available Final ELAIS Catalogue (Rowan-Robinson et
al. \cite{Row04}).
In this paper we consider the surface densities of EROs resulting from 
a wide-field relatively shallow $J$ and $K$-band survey.  We shall 
present the results from a deeper NIR survey around faint mid-IR
detections in subsequent papers (V\"ais\"anen \& Johansson \cite{Va04};
Johansson, V\"ais\"anen \& Vaccari 2004, in prep.). 

The structure 
of this paper is as follows. The observational data are presented in 
\S 2, and in \S 3 we define and extract the ERO sample from the
photometric catalogues.  In \S4 we discuss the dusty vs.\ evolved ERO
separation and in \S5 the SEDs of the detected EROs.  The resulting
surface densities of EROs are presented in \S6.
We assume throughout this paper a flat ($\Omega_{0}=1$) cosmology with
$\Omega_{m}=0.3$, $\Omega_{\Lambda}=0.7$ and $H_{0}=70 \ \rm km s^{-1}
Mpc^{-1}$.

\section{Observations}

\subsection{Near-Infrared data}

The observations presented here were conducted in the ELAIS 
fields N1 and 
N2, centered at $(\alpha,\delta)=16\rm h09\rm m00\rm s,
54^{\circ}40'00\arcsec$ and $(\alpha,\delta)=16\rm h36\rm m00\rm s,
41^{\circ}06'00\arcsec$ J2000.0 respectively.   

The near-IR survey performed with the Mt.Hopkins 1.2-m telescope,
reaching limiting magnitudes of approximately $J=19.3$ 
and  $K=17.5$, is fully presented in V\"ais\"anen et al.\
(\cite{Va00}; hereafter V00).  
The data cover a total of one square degree, although
a slightly smaller area is used here (2850 arcmin$^2$) after some edges
and higher noise areas were excluded and, more importantly, only areas
with full coverage in both bands were considered.  The detector used
was a $256\times256$ InSb array with 1.2\arcsec \, pixels.

\subsection{Optical data} 
\label{opticaldata}

We obtained optical photometric data from the Isaac Newton Telescope
(INT) Wide Field Survey (WFS). 
The WFS data are publicly available on the Cambridge Astronomical
Survey Unit (CASU)  
homepage\footnote{http://archive.ast.cam.ac.uk/}. For a review on the
INT Wide Field  
Survey Project and instrument characteristics, see McMahon et
al. (\cite{Mc01}).  
The data consist of $U,g',r',i',Z$ band photometry (Vega-based) in the
N1 region and $g',r',i',Z$ photometry in the N2 region (the
apostrophes are dropped henceforward for clarity). The WFS bandpasses
are similar to the SDSS (Sloan Digital Sky Survey) filters (Fukugita
et al. \cite{Fu96}).  The WFS 
webpage\footnote{http://www.ast.cam.ac.uk/$\sim$wfcsur/index.php} gives
colour transformations between the WFS filters and those of the standard
Johnson-Cousins system (Landolt \cite{La92}): $r-R = 0.275 (R-I) +
0.008$ and $i-I=+0.211(R-I)$ for the bands most used in this paper.
The nominal $5\sigma$ detection limits for a $1\arcsec$ seeing are
$g\approx25.0$,  $r\approx 24.1$, 
$i\approx23.2$, and $Z\approx22.0$ (the WFS page). 
For further details on the 
pipeline processing of INT wide field survey data consult 
Irwin \& Lewis (\cite{Ir01}), and Gonzalez-Solares et
al. (\cite{Go04}) for more in-depth discussion of
the WFS data in particular in ELAIS fields.
We also obtained raw $R$-band images of the ELAIS N1 and N2 regions from
the archives. 
After reduction we used them for object identification -- only the
pipeline processed catalogues were used for photometry however.

\subsection{Astrometry and photometry}

The INT astrometry was derived using Guide Star Catalog (GSC)
stars, which results in an external 
astrometric accuracy of $0.5\arcsec-1.0\arcsec$ (Irwin \& Lewis
\cite{Ir01}).  Astrometric accuracy in the Mt.Hopkins data ranges between
$0.5-1.5\arcsec$, and was calibrated using GSC and US Naval
Observatory (USNO) catalogues.

All our near-IR photometry is performed using the SExtractor software
(v.2.2.1 and v.2.3; Bertin \& Arnouts \cite{Be96}). 
Our NIR photometry is explained in V00.
As discussed therein, we used total magnitudes as given by the
SExtractor \lq BEST'-magnitude. However,  since the optical WFS
photometry was given only in 2.4\arcsec \,  
diameter aperture magnitudes, we re-did all the old photometry with
matching apertures.  It is crucial that the different 
observations map the same region of the source when constructing
colour indices; near-IR magnitudes {\em not} associated with colours are
given as total magnitudes in this paper.

The archive optical INT WFS data in the ELAIS 
regions come as fully calibrated photometric source catalogues (Irwin
\& Lewis \cite{Ir01}; Gonzalez-Solares et 
al. \cite{Go04}). No further processing is done for the catalogue
except merging of multiple and/or nearby sources, as described below
in Section~\ref{matching}.

\section{Construction of the ERO samples}
\label{ero-samples}

\subsection{Definition of EROs}
\label{definition}

Numerous different selection criteria have been defined for EROs,
including  $R-K\ge 6$, $R-K\ge 5.3$, $R-K\ge 5$, $I-K\ge 4$ 
with $K$-magnitude upper limits from 18 to 21 mag.  All these criteria
are designed for selecting early type galaxies at $z\ge 1$. 
In this paper we use the following definition for EROs: $r-K\ge 5.5$
and/or $i-K\ge 4.4$.  Our limits result from colour
transformations given above in Section~\ref{opticaldata} for typical
colours of $R-I\approx1.4-2.0$ of our sources, and from the
desire to have limits corresponding to the commonly used $R-K> 5$,
$I-K> 4$ selections to aid comparisons to other surveys.  In addition,
we will check any results 
with $r-K>5.8$, i.e.\ $R-K\ge 5.3$, which is also often used. 

Furthermore, we calculated $r-R$ and $i-I$ colours using SEDs from the
GRASIL library (Silva et al. \cite{Si98}) and verified the
$r-R\approx0.5$ and $i-I\approx0.4$ colours to match well those of
ellipticals at $z\sim1$.  Naturally an exact comparison or
transformation between different galaxy colour selection criteria is
redshift and galaxy SED dependent, and we note that
since the model colours that different authors use vary, 
there might easily be 0.1--0.3 mag differences in the colours at $z\sim1$.
Elliptical galaxies become EROs at $z\approx1.1-1.2$ with the criteria
and models we used.

\begin{figure}
\resizebox{8.0cm}{!}{\includegraphics{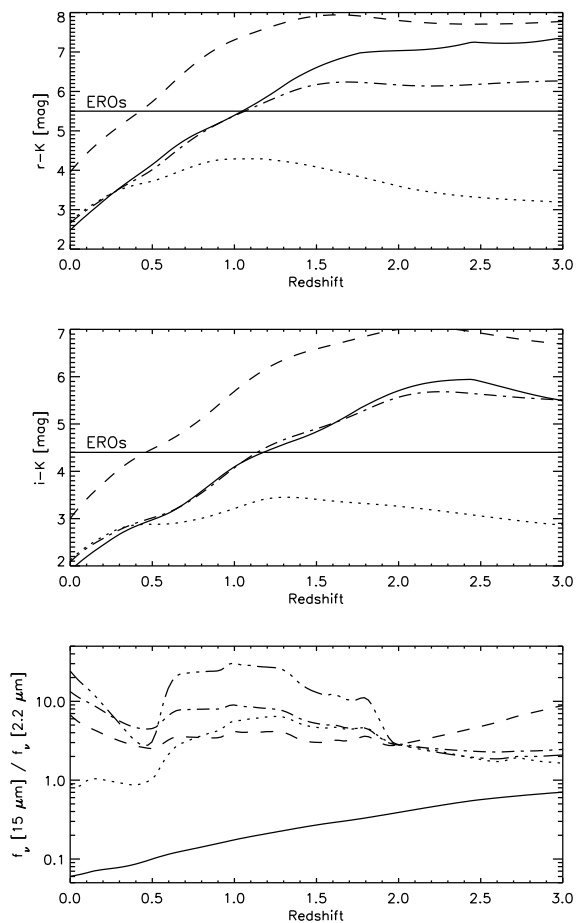}}
      \caption[]{Observed $r-K$ and $i-K$ and MIR/NIR colours of 
representative 
galaxies are calculated from model SEDs of GRASIL code (Silva et al. 
\cite{Si98}). An 
evolving Sb spiral (dotted line) is plotted along with an evolving
elliptical (solid curve). Non-evolving models fitted to the observed
rest-frame SEDs of 
the prototypical starburst M82 (dash-dot) and HR10 ERO (dashed), are
also plotted. Solid horizontal lines show the colour definition of
EROs in this  
paper.  The dash-triple-dot curve in the lowest panel corresponds to the 
ULIRG ARP220, which gives the most extreme MIR/NIR colours.} 
         \label{Silva_model}
   \end{figure}

Fig. \ref{Silva_model} shows $r-K$ and $i-K$ model colours of several 
representative galaxies against redshift with the ERO criteria included.
Model SEDs are adopted from the GRASIL library 
(Silva et al. \cite{Si98}; model SEDs and the GRASIL code are available 
online\footnote{http://web.pd.astro.it/granato/}).
Ordinary 
spirals (dotted line) never reach the red colours of EROs, while both 
ellipticals (solid) and reddened starbursts (dash-dot) become EROs
when seen
beyond $z\sim1$.  For comparison, the colour of the prototype 
dusty ERO HR10 is  
also plotted (dashed curve) as a function of redshift. The colour of the 
HR10 model is due to extreme dust extinction (Silva \cite{Si99}). 
The lowest panel 
shows the flux ratio $f_{15 \mu m}/f_{2.2 \mu m}$ -- 
the degeneracy in the red colours of old 
ellipticals and dusty starbursts is clearly broken, as will be
discussed later in more detail.

\subsection{Matching of optical and NIR data}
\label{matching}

The optical counterparts of near-IR sources in the $J$ and $K$-band
matched Mt.Hopkins catalogue were extracted from the
INT WFS catalogue using 3\arcsec \, search radius.
More than 80\% of the matches are found within
$\approx1.5$\arcsec \, separations, which is reasonable given the
astrometric accuracies and pixel sizes. 
Approximately 4\% of the 6600 NIR sources in the Mt.Hopkins
catalogue were not matched with an optical source.  We checked these
individually: the great majority are close to WFS CCD edges or bright
stars, or were some remaining NIR frame cosmic rays, etc., and though
a handful of these may be genuine very red objects (though not
necessarily EROs) we decided to conservatively exclude all of these
from the final source list.

Since the WFS
catalogue contains multiple detections of the same source (due to 
overlapping CCD frames), optical counterparts within 0.5\arcsec \, of
each other were averaged.  After this purging, 
the remaining multiple matches within 1.5\arcsec \, were 
summed up since the corresponding NIR catalogue would not have
resolved them as individual sources, and then the 
brightest source was selected if there still were multiple
counterparts available.  Approximately 18\% of the original WFS
optical catalogue closest matches were affected by the purging, though
less than 4\% by more than 0.1 magnitudes.

\subsection{Star vs.\ galaxy separation}

We then proceeded to separating
stars from galaxies in the surveys.  Stellarity indices were available
from both the NIR catalogue (the SExtractor CLASS parameter) and the
WFS catalogue, where a flag defines galaxies, definite stars, and
various degrees of uncertain stellarities. The WFS classification is
in principle more useful for us here, since it goes deeper than the
NIR data.  
Note that the star-galaxy separation is different from 
that conducted in V00 since we now have a much deeper optical
catalogue available. 

\begin{figure*}
\resizebox{18.0cm}{!}{\includegraphics{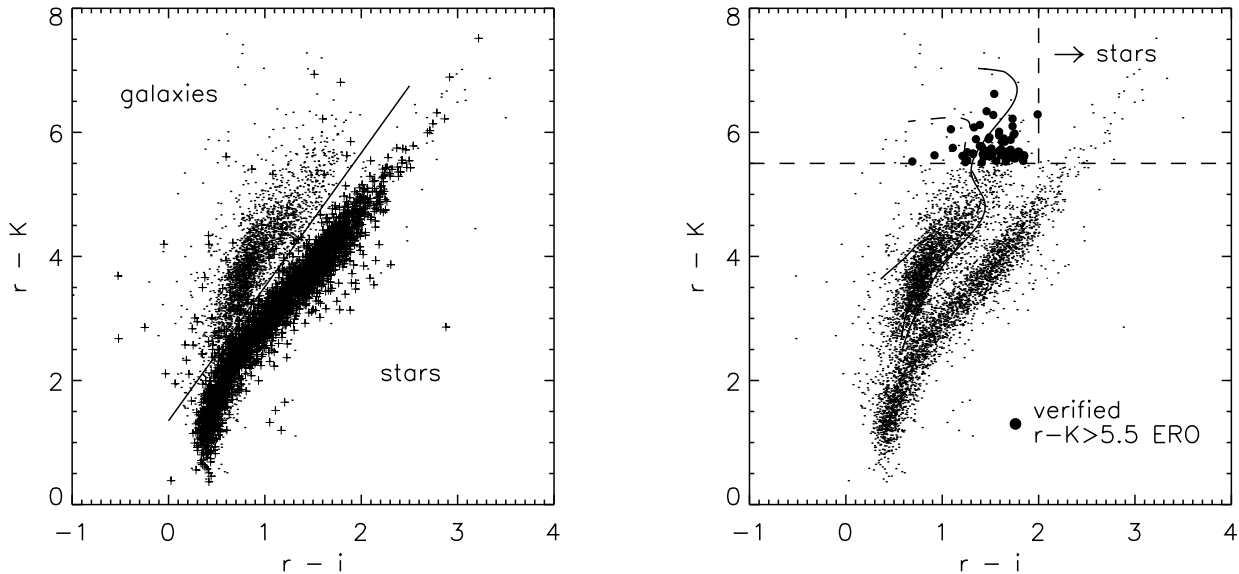}}
      \caption[]{Observed $r-i$ vs.\ $r-K$ colours of our survey.
Stellar sources, as defined by the WFS
classification are overplotted as small crosses
in the left panel.  It is seen that this 
colour-colour diagramme does a very good job in separating stars from
galaxies: in particular there is virtually no overlap in the region of
interest, at $r-K>5$.  Any ``ERO'' with $r-i>2$ is
classified as a star.  The right
panel shows all the galaxy EROs of the Mt.Hopkins survey 
selected with $r-K>5.5$ and several GRASIL
model SEDs overplotted: the lowest solid curve is the SED of
an Sb galaxy,  the dashed curve reaching ERO regime is that of a
starburst, and the highest curve an elliptical.  The curves are
plotted up to redshift of $z=2$ (see text and Sect.~\ref{definition}).
}
        \label{colorfig}
  \end{figure*}

However, for the EROs we are interested in,
the reddest and faintest objects in the catalogue, we find that a colour
separation works best.  This can be seen in Fig.~\ref{colorfig}, which
shows the full catalogue in $r-i$ vs.\ $r-K$ with the
morphological classification indicated: stars are overplotted as small
crosses in the left panel.  A separating line, adjusted
experimentally to maximally distinguish the main concentrations of
stellar and extended sources in the
figure, of $r-K = 2.16(r-i)+1.35$ is drawn.  This colour-colour
diagramme separates stars very well from galaxies; it is important to
note that there is virtually no overlap 
between the stellar sources and galaxies in the region of interest at
$r-K>5$.  
The $r-i$ colour, in fact, is very important in the separation of red
stars. 
Very low mass and cool stars (of the L spectral type in particular)
have $R-K$ and $J-K$ colours closely mimicking those of extragalactic
EROs -- for these stars, however, $R-I$ (or $r-i$) always stays
above $\approx2$ (see e.g.\ Chabrier et al. \cite{Ch00}; Cruz et al.\
\cite{Cr03}).  On the other hand, we find no galaxy models resulting
in $r-i>2$ colours (GRASIL code used; see also eg.\ Fugukita et
al. \cite{Fu95}). All extremely red objects with $r-i>2$ can safely
be discarded as stars. 
  
Representative galaxy 
colours calculated from GRASIL models (Silva et al. \cite{Si98}) are
overplotted in the right panel. 
In fact, the elliptical model does overlap with the $r-i\sim3$
extremely red stars, but only at $z>4$.  This is a potential concern
with deeper ERO surveys, but at our brighter magnitudes such distant
sources are not expected to be seen.

We thus use a combination of
methods.  We define as stellar all 
objects having stellar colours according to the above limit {\em and}
those {\em not} having a galaxy morphology set by the WFS survey. The
end result of this is that brighter sources are preferentially
separated by morphology and fainter ones (especially red sources) by
colour.  The star counts were already compared to the SKY
model predictions of Cohen (\cite{Co94}) in V00, and found to fit
well model predictions for the 
corresponding fields.  In the regime $K<17.5$,
45\% of $r-K>5.5$ sources were classified as stars. 
In summary, it should be stressed that in the case of bright
surveys of EROs, the contamination from red late type (typically
L-type) stars is considerable when using the $R-K$ based selections of
EROs.

\subsection{Extraction of EROs}

We searched for EROs from the Mt.Hopkins 
galaxy catalogue
according to the colour definitions given above in
Section~\ref{definition}. Only NIR detections at $5\sigma$ level and
over were considered.  We then went 
through the resulting list checking our NIR maps as well as the WFS CCD
images for any obvious spurious objects (for example, anomalously low
optical magnitudes were found in cases when sources fell close to gaps
in the CCD frames, or near bright stars -- these were excluded).
Ultimately there were 50 EROs in the matched Mt.Hopkins 
catalogue using the $r-K$ criterion and 21 using the $i-K$ limit - 17
EROs are common to lists resulting from both selection criteria.  
This makes 54 EROs in total.
All the EROs have an $i$-band detection, while there were 4
EROs which have only upper limits in the $r$-band.  
It is obvious that the $r$ and $i$ band based selection criteria for
EROs are not equivalent.  There are more than twice as
much $r-K>5.5$ selected EROs than $i-K>4.4$ selected ones. 

\begin{figure*}
\resizebox{18cm}{!}{\includegraphics{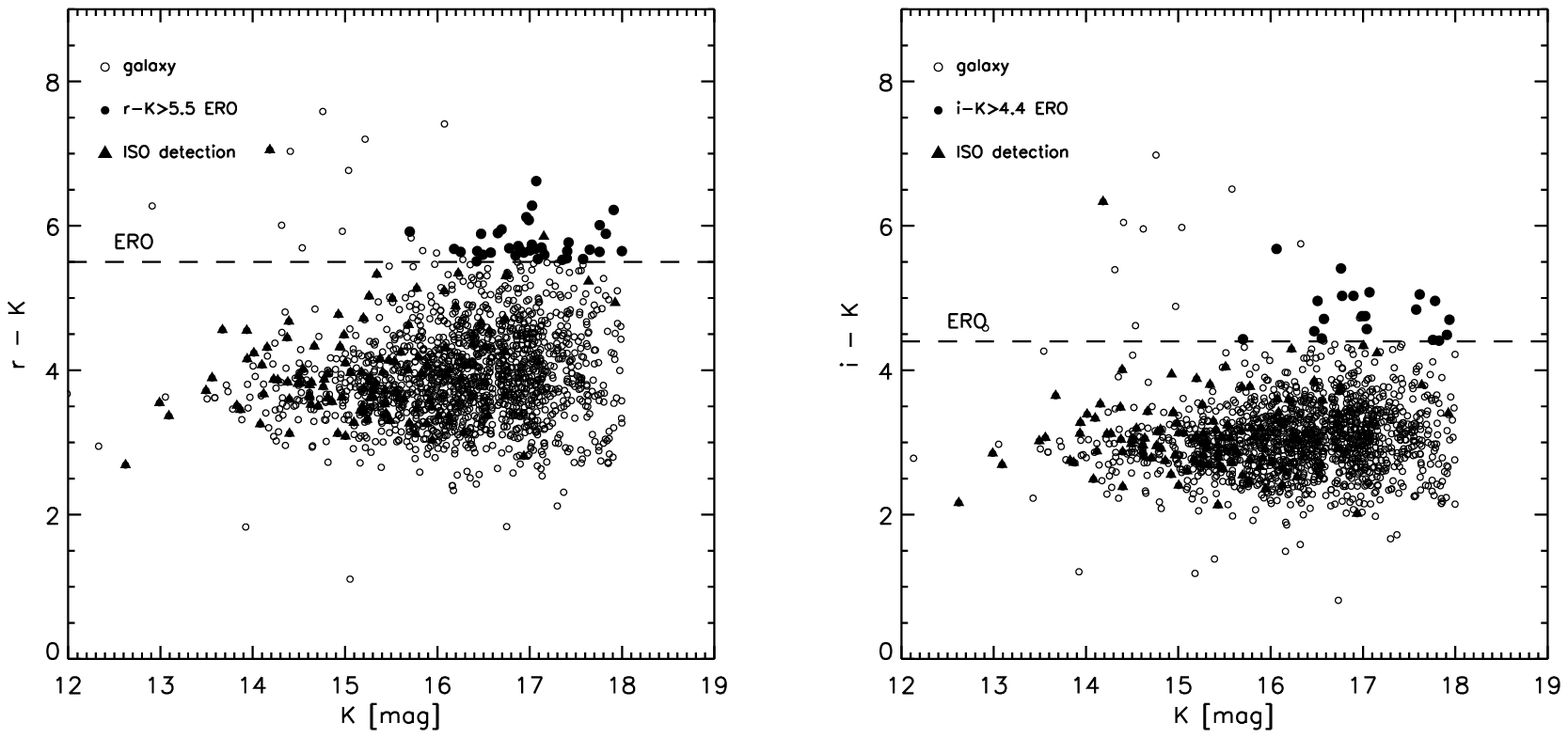}}
      \caption[]{The $K$-band magnitude vs.\ the $r-K$  and $i-K$
	colours, in left and right panels, respectively. All objects
	coming through the catalogue construction as galaxies are
	plotted as open circles.  Individually verified EROs at
	$>5\sigma$ level are
overlaid with solid circles.  Since the survey was done in ELAIS
	fields, we also overplot all matched ISOCAM detections as
	solid triangles.  Note that the total $K$-magnitude is
used on the x-axis, whereas the colour is calculated with matching
small apertures.}
         \label{mth_comb}
   \end{figure*}

The photometry for all EROs is given in Table~\ref{Tab0-phot} and 
the resulting total numbers of verified EROs are summarized in
Table~\ref{Tab1}, along 
with some other survey characteristics.
The colour-magnitude plot is shown in Fig.~\ref{mth_comb}.

We wish to point out, that in order to be very conservative in limiting 
the number of spurious detections in final ERO lists, as well as to be
able to take advantage of the $J-K$ colour in classifiying EROs, we
required both a $J$ and a $K$ detection for all considered objects
from the survey.  While the $5\sigma$ detection requirement was
applied only to the $K$-band, this nevertheless might exclude some
genuine EROs which are very red in $J-K$ colour.  At our faintest
levels of $K\geq17.5$, we do not expect to detect any $J-K\geq2.5$
EROs, and already at $K\geq17$ the completeness of $J-K\geq2$ EROs
would not be quite as high as $J-K<2$ EROs.

  \begin{table*}[b]
\scriptsize
\begin{center}
\begin{tabular}{cccrrrrrr}
     \hline
     \hline
     \noalign{\smallskip}
  &  RA (J2000) & DEC (J2000) & $g'$ & $r'$ & $i'$ & $Z$ & $J$ & $K$  \\
            \noalign{\smallskip}
            \hline
            \noalign{\smallskip}
1 &  16$^h$08$^m$57.0$^s$  &54\degr  43\arcmin  12.2\arcsec & $24.34 \pm  0.12$ & $23.62 \pm  0.14$ & $22.16 \pm  0.09$ & $20.52 \pm  0.05$ & $18.44 \pm  0.13$ & $17.28 \pm  0.20$  \\
2 &  16$^h$08$^m$59.8$^s$  &54\degr  57\arcmin  11.2\arcsec   & $ >25.00      $ & $22.53 \pm  0.05$ & $21.18 \pm  0.03$ & $20.01 \pm  0.04$ & $18.59 \pm  0.19$ & $16.64 \pm  0.14$  \\
3 &  16$^h$09$^m$10.5$^s$  &54\degr  56\arcmin  37.8\arcsec &   $     >25.00  $ & $23.65 \pm  0.14$ & $22.56 \pm  0.10$ & $21.08 \pm  0.10$ & $19.46 \pm  0.27$ &$ 17.60 \pm 0.26$  \\
4 &  16$^h$10$^m$26.7$^s$  &55\degr  04\arcmin  07.5\arcsec & $24.80 \pm  0.16$ & $23.23 \pm  0.11$ & $21.90 \pm  0.05$ & $20.52 \pm  0.06$ & $18.96 \pm  0.20$ & $17.15 \pm  0.24$ \\
5 &  16$^h$10$^m$33.8$^s$  &55\degr  09\arcmin  33.2\arcsec & $      >25.00   $ & $      >24.10   $ & $22.36 \pm  0.08$ & $20.92 \pm  0.00$ & $18.92 \pm  0.27$ & $17.40 \pm  0.22$ \\
6 &  16$^h$10$^m$41.2$^s$  &54\degr  13\arcmin  31.0\arcsec & $       >25.00  $ & $23.07 \pm  0.09$ & $21.96 \pm  0.07$ & $20.90 \pm  0.07$ & $19.50 \pm  0.35$ & $17.32 \pm  0.22$ \\
7 &  16$^h$10$^m$42.8$^s$  &55\degr  01\arcmin  26.8\arcsec & $       >25.00  $ & $23.33 \pm  0.12$ & $21.94 \pm  0.05$ & $20.49 \pm  0.06$ & $19.07 \pm  0.25$ & $17.21 \pm  0.21$ \\
8 &  16$^h$10$^m$45.2$^s$  &55\degr  04\arcmin  50.1\arcsec & $23.84 \pm  0.07$ & $22.34 \pm  0.05$ & $21.42 \pm  0.03$ & $20.60 \pm  0.06$ & $18.89 \pm  0.24$ & $16.71 \pm  0.15$ \\
9 &  16$^h$11$^m$19.1$^s$  &55\degr  07\arcmin  46.5\arcsec & $       >25.00  $ & $23.57 \pm  0.14$ & $22.28 \pm  0.07$ & $21.22 \pm  0.10$ & $19.56 \pm  0.33$ & $17.25 \pm  0.23$ \\
10&  16$^h$12$^m$15.3$^s$  &54\degr  53\arcmin  38.0\arcsec & $       >25.00  $ & $23.96 \pm  0.00$ & $22.42 \pm  0.00$ & $20.93 \pm  0.09$ & $18.92 \pm  0.27$ & $17.34 \pm  0.24$ \\
11&  16$^h$35$^m$12.2$^s$  &40\degr  48\arcmin  35.9\arcsec & $24.58 \pm  0.15$ & $22.63 \pm  0.06$ & $21.14 \pm  0.04$ & $19.68 \pm  0.04$ & $18.67 \pm  0.27$ & $16.71 \pm  0.16$ \\
12&  16$^h$34$^m$45.9$^s$  &40\degr  48\arcmin  26.2\arcsec & $       >25.00  $ & $23.58 \pm  0.14$ & $22.10 \pm  0.10$ & $20.90 \pm  0.08$ & $19.44 \pm  0.27$ & $17.69 \pm  0.25$ \\
13&  16$^h$36$^m$17.9$^s$  &41\degr  15\arcmin  39.8\arcsec & $       >25.00  $ & $23.40 \pm  0.13$ & $21.87 \pm  0.09$ & $20.78 \pm  0.10$ & $19.09 \pm  0.24$ & $17.12 \pm  0.16$ \\
14&  16$^h$36$^m$52.8$^s$  &40\degr  53\arcmin  55.8\arcsec & $       >25.00  $ & $      >24.10   $ & $22.39 \pm  0.00$ & $20.82 \pm  0.07$ & $18.63 \pm  0.18$ & $16.71 \pm  0.14$ \\
15&  16$^h$37$^m$39.6$^s$  &40\degr  53\arcmin  17.0\arcsec & $       >25.00  $ & $23.65 \pm  0.13$ & $21.92 \pm  0.08$ & $21.49 \pm  0.11$ & $19.47 \pm  0.31$ & $17.43 \pm  0.25$ \\
16&  16$^h$38$^m$01.3$^s$  &40\degr  56\arcmin  59.6\arcsec & $       >25.00  $ & $      >24.10   $ & $22.36 \pm  0.00$ & $19.52 \pm  0.02$ & $18.69 \pm  0.22$ & $16.95 \pm  0.16$ \\
17&  16$^h$38$^m$33.7$^s$  &41\degr  07\arcmin  52.0\arcsec & $       >25.00  $ & $      >24.10   $ & $22.44 \pm  0.13$ & $21.07 \pm  0.00$ & $19.61 \pm  0.30$ & $17.41 \pm  0.24$ \\
            \noalign{\smallskip}
\hline
            \noalign{\smallskip}
18&  16$^h$06$^m$52.7$^s$  &54\degr  46\arcmin  37.3\arcsec & $       >25.00  $ & $22.49 \pm  0.05$ & $21.06 \pm  0.03$ & $20.29 \pm  0.03$ & $18.75 \pm  0.19$ & $16.89 \pm  0.20$ \\
19&  16$^h$08$^m$15.6$^s$  &54\degr  41\arcmin  51.7\arcsec & $       >25.00  $ & $22.60 \pm  0.06$ & $20.94 \pm  0.03$ & $20.02 \pm  0.03$ & $19.03 \pm  0.23$ & $17.07 \pm  0.17$ \\
20&  16$^h$08$^m$38.7$^s$  &55\degr  00\arcmin  19.5\arcsec & $       >25.00  $ & $23.14 \pm  0.09$ & $21.52 \pm  0.04$ & $20.45 \pm  0.06$ & $18.88 \pm  0.24$ & $17.29 \pm  0.26$ \\
21&  16$^h$08$^m$59.4$^s$  &54\degr  56\arcmin  34.1\arcsec & $24.72 \pm  0.17$ & $22.84 \pm  0.07$ & $21.33 \pm  0.03$ & $20.08 \pm  0.04$ & $19.06 \pm  0.20$ & $17.10 \pm  0.18$ \\
22&  16$^h$08$^m$60.0$^s$  &54\degr  56\arcmin  25.7\arcsec & $       >25.00  $ & $22.84 \pm  0.07$ & $21.41 \pm  0.04$ & $20.17 \pm  0.05$ & $19.16 \pm  0.23$ & $17.12 \pm  0.19$ \\
23&  16$^h$10$^m$21.0$^s$  &55\degr  07\arcmin  28.1\arcsec & $24.72 \pm  0.15$ & $22.86 \pm  0.08$ & $21.14 \pm  0.03$ & $20.32 \pm  0.05$ & $18.83 \pm  0.17$ & $17.24 \pm  0.24$ \\
24&  16$^h$10$^m$30.3$^s$  &55\degr  08\arcmin  21.1\arcsec & $24.58 \pm  0.13$ & $22.73 \pm  0.07$ & $21.47 \pm  0.03$ & $20.99 \pm  0.09$ & $18.83 \pm  0.19$ & $17.18 \pm  0.20$ \\
25&  16$^h$10$^m$40.4$^s$  &55\degr  05\arcmin  25.7\arcsec & $24.31 \pm  0.11$ & $22.72 \pm  0.07$ & $21.16 \pm  0.03$ & $20.20 \pm  0.04$ & $18.88 \pm  0.23$ & $17.18 \pm  0.21$ \\
26&  16$^h$10$^m$51.0$^s$  &55\degr  12\arcmin  05.8\arcsec & $24.29 \pm  0.11$ & $22.99 \pm  0.08$ & $21.75 \pm  0.04$ & $21.04 \pm  0.09$ & $19.31 \pm  0.33$ & $17.47 \pm  0.24$ \\
27&  16$^h$10$^m$59.1$^s$  &54\degr  31\arcmin  26.2\arcsec & $       >25.00  $ & $22.90 \pm  0.07$ & $21.28 \pm  0.04$ & $20.15 \pm  0.03$ & $19.00 \pm  0.55$ & $17.25 \pm  0.23$ \\
28&  16$^h$11$^m$08.9$^s$  &54\degr  50\arcmin  17.7\arcsec & $       >25.00  $ & $23.07 \pm  0.07$ & $21.64 \pm  0.05$ & $20.85 \pm  0.06$ & $19.42 \pm  0.31$ & $17.42 \pm  0.24$ \\
29&  16$^h$11$^m$12.8$^s$  &55\degr  08\arcmin  24.7\arcsec & $23.43 \pm  0.05$ & $22.57 \pm  0.06$ & $21.25 \pm  0.03$ & $20.17 \pm  0.04$ & $18.73 \pm  0.24$ & $16.91 \pm  0.17$ \\
30&  16$^h$11$^m$14.7$^s$  &55\degr  05\arcmin  31.8\arcsec & $       >25.00  $ & $23.02 \pm  0.09$ & $21.43 \pm  0.03$ & $20.75 \pm  0.07$ & $19.09 \pm  0.26$ & $17.07 \pm  0.21$ \\
31&  16$^h$11$^m$58.4$^s$  &54\degr  53\arcmin  56.5\arcsec & $       >25.00  $ & $22.91 \pm  0.08$ & $21.50 \pm  0.04$ & $20.50 \pm  0.05$ & $18.93 \pm  0.24$ & $17.40 \pm  0.23$ \\
32&  16$^h$34$^m$14.4$^s$  &41\degr  14\arcmin  42.5\arcsec & $24.49 \pm  0.16$ & $22.32 \pm  0.04$ & $20.71 \pm  0.03$ & $19.91 \pm  0.03$ & $18.81 \pm  0.27$ & $16.59 \pm  0.26$ \\
33&  16$^h$34$^m$22.8$^s$  &40\degr  56\arcmin  07.1\arcsec & $       >25.00  $ & $22.76 \pm  0.06$ & $20.96 \pm  0.04$ & $20.23 \pm  0.04$ & $19.13 \pm  0.26$ & $17.12 \pm  0.23$ \\
34&  16$^h$34$^m$52.1$^s$  &40\degr  50\arcmin  50.2\arcsec & $       >25.00  $ & $23.02 \pm  0.08$ & $21.41 \pm  0.05$ & $20.57 \pm  0.05$ & $19.05 \pm  0.23$ & $17.39 \pm  0.20$ \\
35&  16$^h$34$^m$58.4$^s$  &40\degr  52\arcmin  55.4\arcsec & $       >25.00  $ & $22.80 \pm  0.06$ & $21.10 \pm  0.04$ & $20.36 \pm  0.04$ & $18.86 \pm  0.19$ & $17.20 \pm  0.16$ \\
36&  16$^h$35$^m$12.1$^s$  &40\degr  47\arcmin  31.0\arcsec & $       >25.00  $ & $23.31 \pm  0.11$ & $21.82 \pm  0.08$ & $20.69 \pm  0.07$ & $19.13 \pm  0.27$ & $17.60 \pm  0.25$ \\
37&  16$^h$36$^m$01.0$^s$  &41\degr  05\arcmin  59.3\arcsec & $       >25.00  $ & $22.99 \pm  0.08$ & $21.58 \pm  0.06$ & $20.30 \pm  0.05$ & $19.01 \pm  0.17$ & $17.22 \pm  0.21$ \\
38&  16$^h$36$^m$07.5$^s$  &41\degr  21\arcmin  42.3\arcsec & $24.69 \pm  0.00$ & $22.39 \pm  0.04$ & $20.97 \pm  0.00$ & $19.95 \pm  0.00$ & $19.11 \pm  0.27$ & $16.74 \pm  0.23$ \\
39&  16$^h$36$^m$07.8$^s$  &41\degr  03\arcmin  41.4\arcsec & $       >25.00  $ & $22.94 \pm  0.07$ & $21.35 \pm  0.07$ & $20.78 \pm  0.07$ & $19.56 \pm  0.27$ & $17.24 \pm  0.23$ \\
40&  16$^h$36$^m$58.1$^s$  &40\degr  49\arcmin  43.9\arcsec & $       >25.00  $ & $22.41 \pm  0.05$ & $20.74 \pm  0.03$ & $19.95 \pm  0.03$ & $18.64 \pm  0.20$ & $16.72 \pm  0.11$ \\
41&  16$^h$37$^m$23.9$^s$  &41\degr  01\arcmin  16.0\arcsec & $       >25.00  $ & $23.27 \pm  0.10$ & $21.42 \pm  0.09$ & $20.66 \pm  0.05$ & $19.19 \pm  0.27$ & $17.64 \pm  0.26$ \\
42&  16$^h$37$^m$27.5$^s$  &40\degr  55\arcmin  43.6\arcsec & $       >25.00  $ & $23.03 \pm  0.00$ & $21.27 \pm  0.00$ & $20.90 \pm  0.08$ & $18.97 \pm  0.24$ & $17.45 \pm  0.26$ \\
43&  16$^h$37$^m$31.0$^s$  &40\degr  53\arcmin  36.3\arcsec & $23.93 \pm  0.10$ & $22.47 \pm  0.05$ & $20.76 \pm  0.03$ & $19.99 \pm  0.03$ & $18.57 \pm  0.20$ & $16.79 \pm  0.16$ \\
44&  16$^h$37$^m$38.6$^s$  &40\degr  56\arcmin  59.3\arcsec & $       >25.00  $ & $23.00 \pm  0.07$ & $21.36 \pm  0.05$ & $20.39 \pm  0.04$ & $19.19 \pm  0.40$ & $17.10 \pm  0.17$ \\
45&  16$^h$37$^m$40.2$^s$  &40\degr  54\arcmin  21.8\arcsec & $       >25.00  $ & $22.64 \pm  0.05$ & $20.80 \pm  0.05$ & $19.66 \pm  0.05$ & $18.86 \pm  0.23$ & $17.10 \pm  0.20$ \\
46&  16$^h$37$^m$49.6$^s$  &40\degr  55\arcmin  43.5\arcsec & $       >25.00  $ & $22.66 \pm  0.05$ & $20.86 \pm  0.03$ & $20.12 \pm  0.03$ & $18.92 \pm  0.24$ & $16.98 \pm  0.17$ \\
47&  16$^h$37$^m$57.2$^s$  &41\degr  19\arcmin  38.4\arcsec & $24.18 \pm 0.00 $ & $23.09 \pm  0.08$ & $21.57 \pm  0.06$ & $20.75 \pm  0.08$ & $19.65 \pm  0.31$ & $17.45 \pm  0.22$ \\
48&  16$^h$37$^m$58.1$^s$  &41\degr  09\arcmin  47.6\arcsec & $24.24 \pm  0.12$ & $22.80 \pm  0.06$ & $21.48 \pm  0.05$ & $20.56 \pm  0.06$ & $19.40 \pm  0.28$ & $17.15 \pm  0.19$ \\
49&  16$^h$38$^m$20.3$^s$  &41\degr  03\arcmin  42.5\arcsec & $23.73 \pm  0.08$ & $22.21 \pm  0.04$ & $20.55 \pm  0.02$ & $19.76 \pm  0.03$ & $18.72 \pm  0.31$ & $16.62 \pm  0.23$ \\
50&  16$^h$38$^m$46.1$^s$  &41\degr  08\arcmin  18.8\arcsec & $       >25.00  $ & $23.37 \pm  0.10$ & $21.64 \pm  0.06$ & $20.48 \pm  0.06$ & $19.39 \pm  0.29$ & $17.27 \pm  0.24$ \\
            \noalign{\smallskip}
\hline
            \noalign{\smallskip}
51&  16$^h$10$^m$35.1$^s$  &55\degr  10\arcmin  04.0\arcsec & $23.55 \pm  0.06$ & $22.69 \pm  0.06$ & $21.89 \pm  0.05$ & $20.92 \pm  0.08$ & $19.36 \pm  0.35$ & $17.47 \pm  0.26$ \\
52&  16$^h$11$^m$13.3$^s$  &54\degr  52\arcmin  48.3\arcsec & $22.86 \pm  0.05$ & $21.99 \pm  0.06$ & $21.34 \pm  0.04$ & $20.18 \pm  0.04$ & $18.65 \pm  0.17$ & $16.94 \pm  0.17$ \\
53&  16$^h$34$^m$21.4$^s$  &40\degr  57\arcmin  56.2\arcsec & $23.96 \pm  0.09$ & $22.96 \pm  0.08$ & $22.19 \pm  0.10$ & $21.21 \pm  0.10$ & $19.35 \pm  0.28$ & $17.50 \pm  0.25$ \\
54&  16$^h$35$^m$46.0$^s$  &41\degr  14\arcmin  38.6\arcsec & $23.66 \pm  0.15$ & $22.64 \pm  0.08$ & $22.02 \pm  0.08$ & $20.98 \pm  0.07$ & $19.76 \pm  0.28$ & $17.18 \pm  0.22$ \\
\noalign{\smallskip}
\hline
\end{tabular}
\end{center}
\caption[]{Photometry of the whole ERO sample.  Sources 1--17 are those
  with  $r-K>5.5$ and $i-K>4.4$, while 18--50 and 51--54 are selected
  by only $r-K>5.5$ and $i-K<4.4$, respectively.}
\label{Tab0-phot}
\end{table*}

   \begin{table*}
\begin{center}
\begin{tabular}{lccccccccccccc}
     \hline
     \hline
     \noalign{\smallskip}
Sample   & N  & $N_{\star}$ & Area $\rm arcmin^{2}$ & $K_{\rm lim}$ & $<r-K>$ & $<i-K>$ & \multicolumn{3}{c}{EROs} & \multicolumn{2}{c}{$\rm ERO/arcmin^{2}$} \\
(1)      & (2) & (3) & (4) & (5) & (6) & (7) & (8) & (9) & (10) & (11)   & (12) \\
            \noalign{\smallskip}
            \hline
            \noalign{\smallskip}
Mt.Hopkins   & 6618 & 3306 & 2850  & 17.5 & 3.91 & 3.09 & 50 & 21 & 17 &  0.007--0.018  & 0.004--0.007  \\
\noalign{\smallskip}
\hline
\end{tabular}
\end{center}
\caption[]{Columns (2) to (5) give the total
  numbers of sources and stellar objects, the surveyed area, and the
  average limiting magnitude. The rest of the columns refer to
  galaxies only: columns (6) to (7) give mean colours, though note
  that these are lower limits since optical non-detections we ignored.
  Column (8) gives the  number of EROs selected by
  $r-K\ge5.5$, (9) by $i-K\ge4.4$, and (10) by their combination; column
  (11) is the range of surface densities resulting from either $r$ or 
$i$ band based selection; column (12) is the range of cumulative 
surface densities at $K<17$ }
\label{Tab1}
\end{table*}

\section{Separation of dusty EROs}
\label{dustysep}

Since the Extremely Red Objects can be divided into two broad classes,
the populations should be differentiated before 
surface density comparisons to detailed galaxy formation models are
attempted. 
A prime motivation of this study is to attempt the separation using
the mid-IR data.  We shall also perform the separation using other
established methods, and discuss the differences.

\subsection{Using mid-IR data}

Since only dusty EROs are expected to show up in the mid-IR ELAIS
survey, the question to answer is, what fraction of the EROs are
detected in the mid-IR {\em to a given flux limit}?   

Among the 54 Mt.Hopkins EROs we find only one match with a 15 
$\mu$m ISOCAM source.  In the ELAIS band-merged catalogue
(Rowan-Robinson et al.\ \cite{Row04}) the source is a candidate 
hyper luminous IR-galaxy based on a photometric redshift of
$z\approx1.0$, which agrees with our independent photometric redshift
determination of $z=0.9$.
(We note that using the same near-IR data-set but with
only $K$-band considered, another ISOCAM detected ERO is
found in Rowan-Robinson et al.\ (\cite{Row04})).
Thus, at face value, the fraction of dusty EROs
seems to be insignicant.  

However, it is crucial to take into account detection limits.
The main hindrance to the full power of the mid-IR separation with our
data-set is the relative shallowness of the ELAIS survey.  
All the Mt.Hopkins EROs have $K=16-18$ mag.  
Comparing to GRASIL models (Fig.~\ref{Silva_model}),
a starburst galaxy with M82-type SED and $K=17$ apparent magnitude
should just have been detected over the 0.7 mJy ISOCAM flux limit. 
In comparison, to detect the Rayleigh-Jeans tail of ellipticals would
have required approximately $K=13$ brightness.  
We note that M82-type galaxy with $K=17$ mag at $z=0.7$ and $z=1.0$
translates to star formation rates $SFR \approx 100$ and 200 ${\rm
M_{\sun} yr^{-1}}$, respectively, as calculated from total IR
luminosity $L_{\rm IR} (3-1000 \mu{\rm m})$ with relation given e.g.\
in Mann et al. (\cite{Ma02}) and using the GRASIL model SED. The
corresponding IR luminosities are $log(L_{\rm IR}) > 11.7$. 

There are 23 $K<17$ EROs in the sample, out of which only one (4\%) 
is detected in the mid-IR.  As shown above, the ELAIS data is only
sensitive to luminous IR galaxies at the expected redshifts of EROs
($z>0.7$).  Thus, we can state that very strong starbursts ($SFR > 100
\, M_{\sun} yr^{-1}$ make up only a small fraction, less than 10\%,
of counterparts to bright EROs.  For detection of more modest dusty
galaxies other methods have to be used.

It is relevant to note that the in-orbit Spitzer mission will
definitely find large numbers of dusty EROs, including more modest
ones than above:  Using the
M82 SED once again as an example, sampling dusty $K\approx20$ EROs
at $z\approx1$ would mean probing dusty starforming galaxies
of $SFR \sim 10 \, {\rm M_{\sun} yr^{-1}}$. The expected flux
densities of such objects with IRAC 8$\mu$m and MIPS 24$\mu$m bands
would be of the order of 10 and 150 $\mu$Jy, respectively, which are
easily reached 
in a few minutes of integration time.  For example, assuming 10 minute
per pixel integrations, a 10 hour survey to the mentioned $5\sigma$
depths would cover $\sim 800$ arcmin$^{2}$ in both bands.  This means
detecting 100--200 MIR counterparts of dusty EROs, assuming average ERO
surface densities and an {\em ad hoc} 50\% fraction of dusty EROs.

\subsection{Colour-colour selections}

\begin{figure*}
\resizebox{18.0cm}{!}{\includegraphics{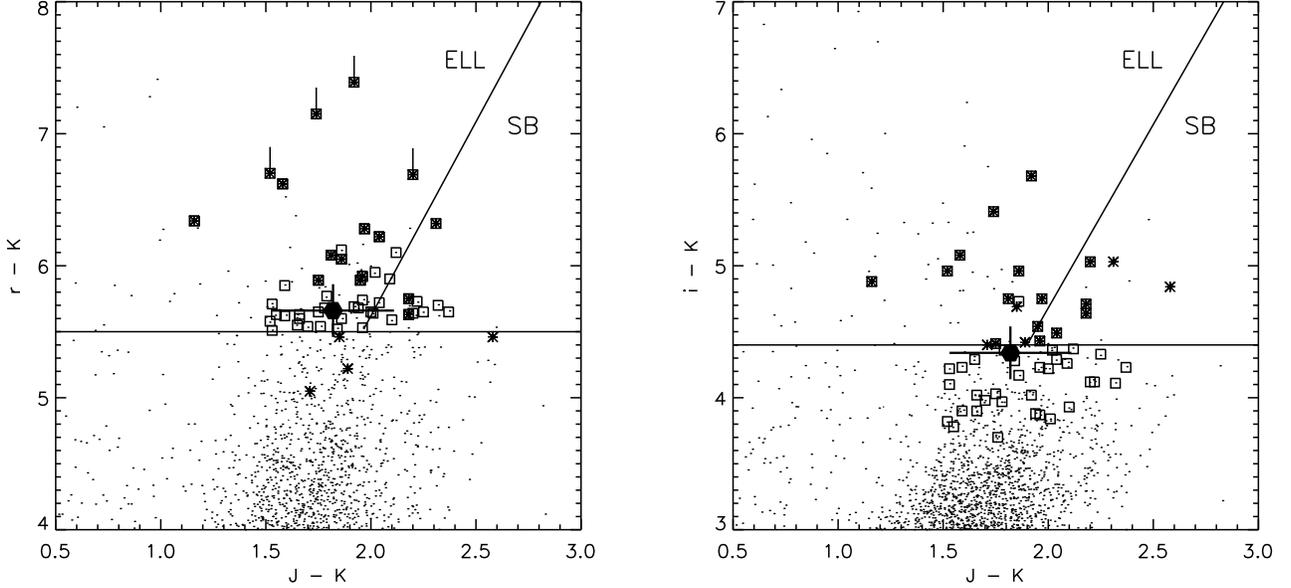}}
      \caption[]{
$J-K$ vs.\ $r-K$ and $i-K$ ERO discriminator colours adopted from
	Pozzetti \& Mannucci (\cite{Po00}).  All 
	galaxies are plotted as dots and EROs are overlaid 
	with larger symbols.  Squares are $r-K>5.5$  EROs and asterisks
	$i-K>4.4$ EROs.
	Those $i-K$ EROs with no $r$-band detection
	have lower limits indicated in the left panel.  The only
	mid-IR detected ERO is shown as the large solid symbol -- its
	error bars are representative to errors of all EROs.}
         \label{jk_sep}
   \end{figure*}

Figure~\ref{jk_sep} shows the colour-colour separation scheme of
EROs by Pozzetti \& Mannucci (\cite{Po00}) adopted for our $r,i$
filters. The idea is 
that the $J-K$ separates the EROs at $z>1$ to bluer early type
galaxies (where the large optical to near-IR colour is due to the
4000\AA \ break), and to the redder (in $J-K$) dusty EROs which have a
a smoother SED from optical to near-IR.  In the $r-K$ plot we find
9/50 $r-K$ selected EROs in the dusty starburst side of the indicator, 
and similarly 3/16 of the $i-K$ EROs.  In the $i-K$ plot many $r-K$
EROs fall under the $i-K=4.4$ line, and the dusty percentages are
higher for the remaining EROs, $30-50$\%.
The one mid-IR ERO is overplotted in the figure, and it in fact
falls on the elliptical side of the division.  However, given the
typical photometric errors of the EROs (the error bars on the mid-IR
ERO are representative) it is seen that at least 1/3 of the EROs fall
statistically on the dividing line.  Moreover, while the elliptical
vs.\ starburst separating line is equivalent to Pozzetti
\& Mannucci, it is only defined to work for $R-K>5.3$.  If we thus
select EROs at $r-K>5.8$, nearly all galaxies on the starburst side of
the left panel in Fig.~\ref{jk_sep} fall out, and the elliptical
fraction becomes close to 90\%. 
Finally, our Mt.Hopkins 
sample is 1-2 magnitudes brighter than other typical ERO samples, and
it is therefore not certain how accurate the separation should be.  

We also checked the colour-colour separation scheme presented
  by Bergstr\"om \& Wiklind (\cite{Be04}) using $R-J$ vs.\ $J-K$
  colours (see their Fig.~8).  The result is 85\% ellipticals
  regardless of whether $r-K$ or $i-K$ ERO criterion (or both) is used,
  totally consistent with the fractions emerging from the Pozzetti \&
  Mannucci $R-K$ vs.\ $J-K$ method. 

We thus conclude that using the various $J-K$ vs.\ optical-NIR colour   
separation schemes approximately 80\% of our EROs appear ellipticals.
Only when using Pozzetti \& Mannucci $i-K$ based separation for
$i-K>4.4$ EROs, is the percentage lower, $\approx 60$\%.
We might be missing the very reddest $J-K$ EROs, i.e.\ some dusty
EROs, because of the selection of the sample in both bands.
At $K<17$, where there should be no bias in $J-K$ colour, the
fractions of early type EROs range between 65-80\%, depending 
on the optical band used.

\section{Photometric redshifts and SEDs}
\label{templates}

\begin{figure}
\resizebox{9.0cm}{!}{\includegraphics{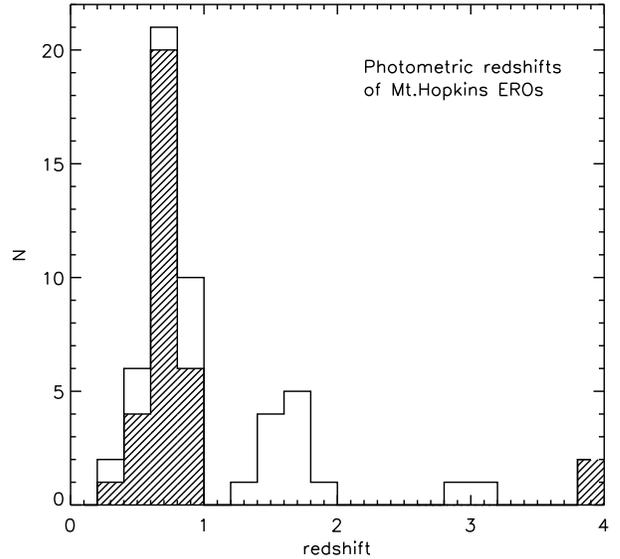}}
      \caption[]{Photometric redshifts as calculated by HYPERZ.
All 54 Mt.Hopkins EROs are plotted, whatever the
confidence of the fit (half have $>85$\%).  $r-K>5.5$ selected EROs which
are {\em not} EROs with $i-K>4.4$ selection are hatched, the empty
regions of the histogram thus show the  $i-K>4.4$ EROs.  }
        \label{zphot}
  \end{figure}

We calculated photometric redshifts for the EROs
using the HYPERZ code (\cite{Bo00}) which fits GISSEL98 synthetic
spectra (Bruzual \& Charlot \cite{Br93}; BC hereafter) galaxy
templates to photometric data points. 
Figure~\ref{zphot} shows the photometric redshift distributions of our
sample.  The mean photometric redshift is $z=0.94 \pm 0.49$,
excluding 
the outliers at $z\sim4$.  However, it is clear that subgroups are 
involved, though the gap at $z=1-1.2$ is difficult to understand.  We
also note that approximately half of the fits result in a confidence
better than $>85$\%, so the results should be taken with some
caution.  On the other hand, the overall distribution, average
redshifts, or properties of sub-groups discussed below do not change
if we consider only those EROs with fits of $>85$\% confidence. 

The 21 $i-K$ selected EROs, the white regions in the
histogram, are more evenly distributed and at higher $z$ (average
$z\approx1.4$) than $r-K$ EROs.  The 17 objects which are EROs with
both selection criteria cover quite evenly the range $z=0.6-1.8$
(average at $z\approx1.3$). The EROs \#51--54 in Table~\ref{Tab0-phot}
are all between $z=1.4-1.9$.  

Most notably, however, those EROs that are {\em 
not} EROs with the $i-K>4.4$ criteria, show a significantly narrower
redshift distribution (hatched region of histrogram) with $z=0.73 \pm
0.06$.  Moreover, these objects constitute more than half, 33/54, of
our total ERO sample -- we return to these below in
Section~\ref{evolved}.  Taken together, the photometric redshifts thus
suggest that $r-K>5.5$ selects more nearby systems at redshift unity
and below, and $i-K>4.4$ includes a wider range of EROs. 

\begin{figure}
\resizebox{9.0cm}{!}{\includegraphics{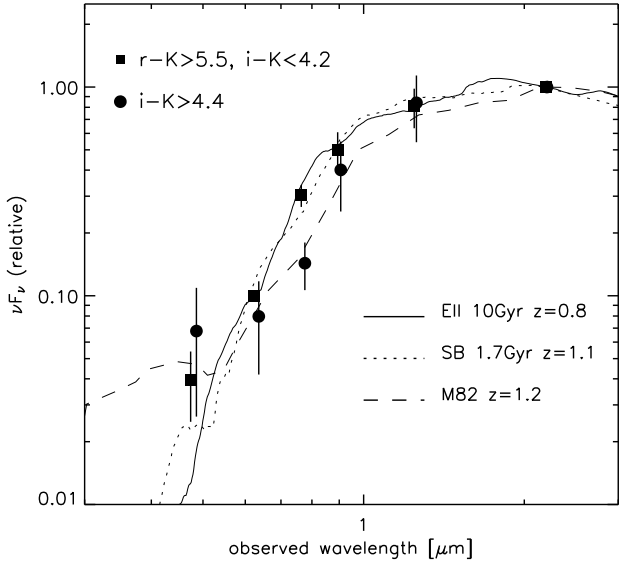}}
      \caption[]{Averaged and $K$-band normalized SEDs of $i-K$ EROs
	(circles) and those $r-K$ EROs with $i-K<4.2$ (squares). The
	observed bands are $g,r,i,Z,J,K$. The GRASIL M82 model and
	elliptical and evolved starburst SEDs 
	from BC are also plotted at indicated redshifts.  
}
         \label{eroseds}
   \end{figure}

Virtually all, 50/54, of the best-fit HYPERZ BC SEDs are
either starburst spectra of age 0.4 to 2 Gyr, or elliptical
spectra at several Gyr -- these SEDs are nearly identical (see
Fig.~\ref{eroseds}).  In fact, in a recent paper by Pierini et
al.\ (\cite{Pi04}) it was proposed that post-starburst ellipticals,
forming in a short bursts {\em during} the period where EROs are
observed (between $1<z<2$), are a neglected constituent of ERO
populations.   It should be stressed that basically all of the
best-fit HYPERZ 'starbursts' are evolved results of
instantenious bursts and {\rm not} dusty starforming
galaxies.

\section{Discussion}

\subsection{ERO counts and surface densities}

\begin{figure*}
\resizebox{18.0cm}{!}{\includegraphics{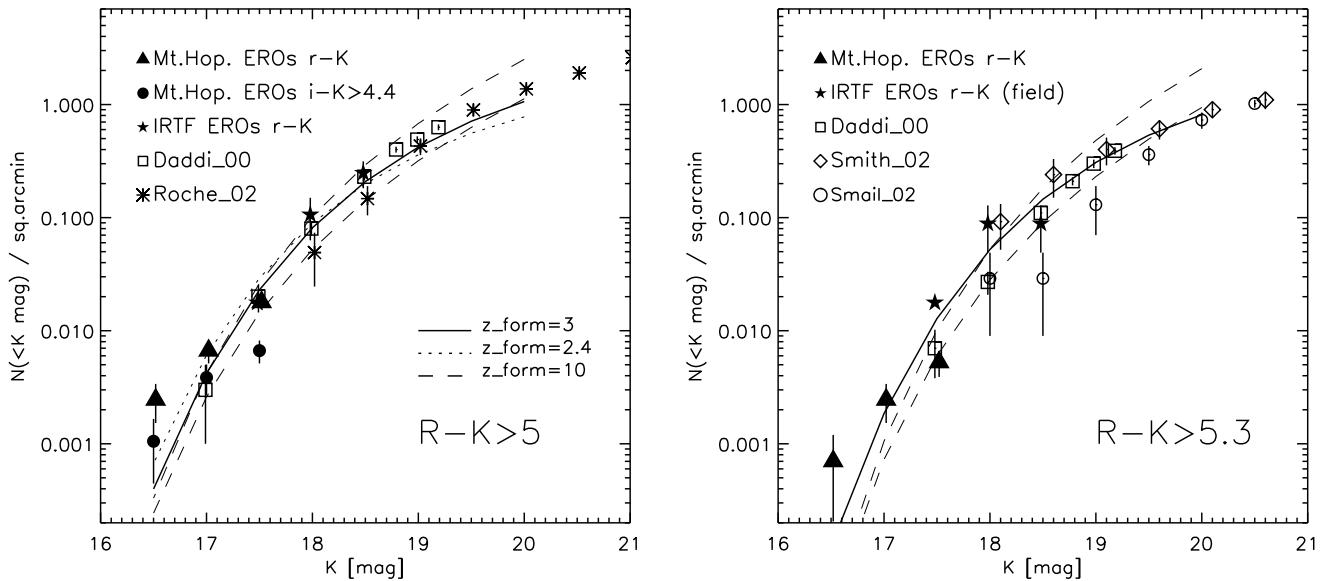}}
      \caption[]{Cumulative ERO counts are shown for $R-K>5$ and
	$R-K>5.3$ selections (ie. $r-K>5.5$ and $r-K>5.8$), in the
	left and right panels, 
	respectively, with solid triangles. Additionally, the
	$i-K>4.4$ selected EROs are 
	plotted in the left panel with circles. Error bars are
	Poissonian. Our other
	field ERO counts from a deeper survey (IRTF, V\"aisanen \&
	Johansson 2004) are shown as stars. Squares show 
	the EROs of Daddi et al.\ (\cite{Da00}) and the asterisks the
	Roche et al.\ (\cite{Ro02}) EROs in the left panel.  The
	left panel shows $R-K>5$ selected EROs of Daddi et
	al. (\cite{Da00}) and Roche et al.\ (\cite{Ro02}).  In
	addition to the Daddi et al.\ $R-K>5.3$ selected EROs the
	right panel plots the corresponding Smith et al.\
	(\cite{Smi02}) and Smail et al.\ (\cite{Sm02}) EROs.
        We plot PLE
	models of Daddi et al.\ calculated using the indicated $R-K$
	cuts. The curves are different by their 
	galaxy formation redshifts, as indicated in the figure.  The
	two dashed-lines employ a different LF -- the higher curve
	results from a 2MASS 
	normalized LF, and the lower from that of Marzke et al.\
	(\cite{Ma94}). 
	}
         \label{allcounts}
   \end{figure*}

The numbers of detected EROs are given in Table \ref{Tab2}, and
the corresponding cumulative number counts are plotted in
Fig.\ref{allcounts}.  The numbers plotted are the total ERO counts:
based on the previous discussion, the fraction of dusty EROs is low
($\sim15$\%) and does not have significant effect even if subtracted
from the counts.

   \begin{table*}
\begin{center}
\begin{tabular}{lcccccccccc}
     \hline
     \hline
     \noalign{\smallskip}
             &   \multicolumn{3}{c}{$r-K>5.5$} & \multicolumn{3}{c}{$r-K>5.8$} & \multicolumn{3}{c}{$i-K>4.4$} \\
            \noalign{\smallskip}
$K$ limit &  & Frac. & $\Sigma_{K}$ & & Frac. & $\Sigma_{K}$ & & Frac. & $\Sigma_{K}$  \\
(mag)     &  N & \% & $\rm arcmin^{-2}$ & N & \% & $\rm arcmin^{-2}$ & N & \% & $\rm arcmin^{-2}$ & \\
            \noalign{\smallskip}
            \hline
            \noalign{\smallskip}

$K \leq 16.5$  & 7 & 0.004 & 0.002 & 2 & 0.001 & 0.0007 & 3 & 0.002 &  0.001 \\
$K \leq 17.0$  & 19 & 0.008 & 0.007 & 7 & 0.003 & 0.002 & 11 & 0.005 &  0.004 \\
$K \leq 17.5$  & 51 (35) & 0.012 & 0.018 & 15 (11) & 0.004 & 0.005 & 19 (15) & 0.005 & 0.007 \\
\noalign{\smallskip}
\hline
\end{tabular}
\end{center}
\caption[]{The sample of Mt.Hopkins EROs.  Cumulative counts are given
for different ERO colour selection criteria.  Total numbers (N),
Fraction of EROs compared to the total galaxy count (``Frac.'') is
calculated using galaxy counts in these same fields (V\"ais\"anen et
al.\cite{Va00}).  At $K\leq17.5$ N shows the completeness corrected
value, and the raw count is given in parentheses.}
\label{Tab2}
\end{table*}

Our wide survey from Mt.Hopkins extends the ERO number counts 
to brighter magnitudes than observed before.  We report the first ERO
counts at $K<16.5$, where the $r-K>5.5$ (ie.\ $R-K>5$) selection
yields a surface density of $0.002 \pm 0.001 \rm arcmin^{2}$, and the
$i-K>4.4$ (ie.\ $I-K>4$)  selection $\approx50$\% of this value.
At $K<17$ magnitudes we arrive at 
$0.007 \pm 0.002 \rm arcmin^{2}$ of $r-K$ EROs.  
Beyond $K>17$ mag the
Mt.Hopkins survey is starting to be affected by incompleteness, and
based on simulations performed in V00
for the same dataset, we estimate a factor of 1.5 correction to the
$K<17.5$ cumulative ERO density.  The resulting $r-K$ selected surface
density at $K<17.5$ is $0.018 \pm 0.003 \rm arcmin^{2}$.

We compare these numbers to the only two other
field ERO surveys with enough sky coverage to reach these bright
magnitude levels: 
The largest ERO survey to date Daddi et al.\ (\cite{Da00}; with 701
arcmin$^{2}$), 
yielded $0.003 \pm  
0.002 \rm arcmin^{2}$ at $K<17$ and $0.02 \pm
0.006 \rm arcmin^{2}$ at $K<17.5$.  These are totally consistent with
our result. 

The Yan \& Thompson (\cite{Ya03}; 409 arcmin$^{2}$) HST ERO counts
were selected 
using $I-K>4.0$.  At $K<17$ they agree with our corresponding
$i-K>4.4$ selection within the errors.  At $K<17.5$ 
there is a clear discrepancy, however.
It is our $r-K$ selected EROs which are in closer agreement with the
Yan \& Thompson  ERO counts -- we systematically find a factor of
$\sim2$ less $i-K$ selected EROs than $r-K$ selected ones.  Especially
at the
faintest bin there appears to either be more incompleteness than we
expect, or the $i-K>4.4$ selection does not pick up as many bright
$K\sim17$ galaxies at $z\sim1$ than the $r-K$ selection does.  Indeed,
the discussion on ``excess'' $r-K$ EROs in the next
Section~\ref{evolved} and the photometric redshifts derived
for the EROs hint at the latter possibility.

\subsection{Evolved ellipticals at $z\sim1$}
\label{evolved}

\begin{figure*}
\resizebox{18.5cm}{!}{\includegraphics{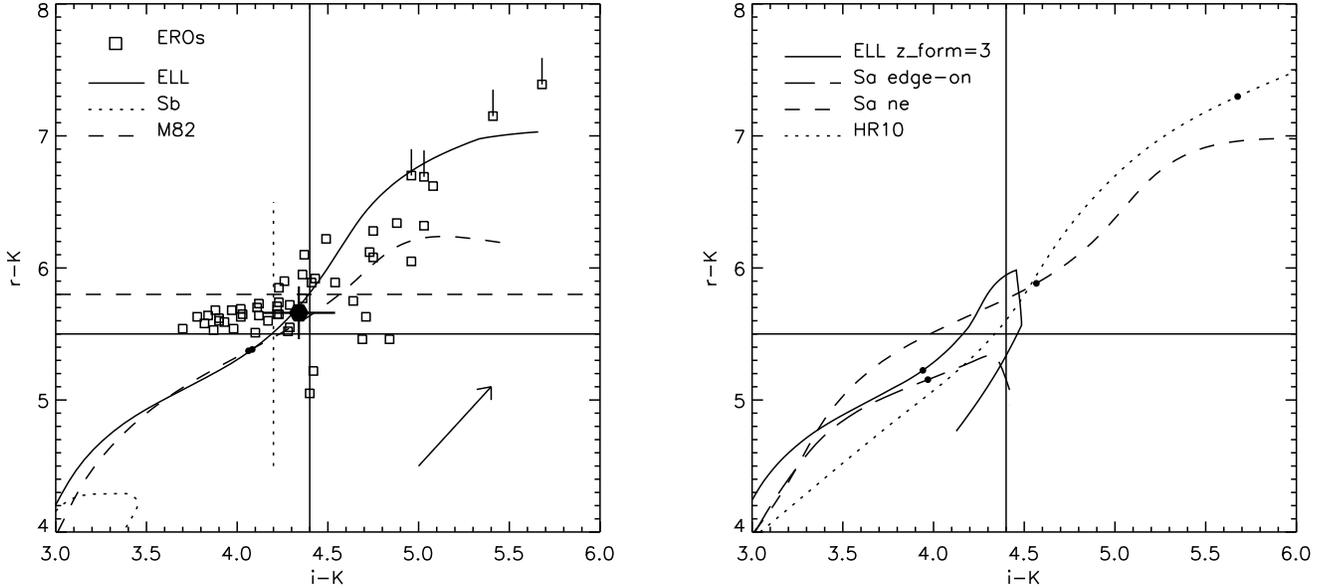}}
      \caption[]{$i-K$ vs.\ $r-K$ colour-colour magnitude diagramme
	of the verified EROs is plotted in the left panel. The
	mid-IR detected EROs is overplotted as a large solid symbol.
Several GRASIL models are plotted in both panels:  the models are drawn
	until redshift $z=2$, and the $z=1$ locations are indicated by
	small solid circles along the curves. The reddening direction
	is indicated in the 
left panel by the diagonal arrow which correcponds to $A_{V}=1$
extinction.  
} 
        \label{ikrk}
  \end{figure*}

The strength, and motivation, of ERO searches has been to look for the
most massive galaxies at redshifts of unity and over (see e.g.\
Saracco et al.\ \cite{Sa03}), since it is exactly these which place
tight constraints on galaxy formation scenarios.  
The dusty vs.\ early type ERO discriminators showed that most of our
ERO sample is consistent with being early type galaxies.  However, we
argue that the clearest population of ellipticals comes from the class
of $r-K>5.5$ EROs which have $r-K<4.4$ -- these constitute 60\% of all
our EROs.  Fig.~\ref{ikrk} shows an $i-K$ vs.\ $r-K$ plot, where these
``excess'' EROs populate the upper right quadrant (see also
Figs.~\ref{jk_sep} and~\ref{eroseds}). 

Intuitively, the upper left quadrant galaxies should be sources at a
redshift where a significant 4000 \AA \ break falls right in
between the $r$ and $i$ bands: i.e.\ ellipticals or very early type
spirals at a redshift $z \sim 0.8$.  
We calculated numerous GRASIL models, and it turns out to be fairly
difficult to obtain colours in this ERO regime.  Any significant
on-going star-formation drops the $r-K$ colour down, as noted also by
e.g.\ Yan \& Thompson (\cite{Ya03}; see their Fig.9.).  The effect is
seen in the solid curve of the right panel of Fig.~\ref{ikrk}: the
elliptical model has a formation redshift of $z_{\star} = 3$ and
at redshifts of $z>1.5$ the effects of the starburst turn
the $r-K$ colour sharply down.  Significant amounts of extinction,
including edge-on spirals, do not bring models to the $r-K>5.5,
i-K<4.4$ region either: the dotted line depicts the extremely dusty
HR10 model as an example.  

One gets closest to the $r-K>5.5, i-K\sim4$ region by merely  
redshifting an old present-day elliptical to Sa galaxy to the
appropriate redshift.  This is shown as the dashed curve, a 
13 Gyr old Sa spiral; any $>10$ Gyr old elliptical produces a similar
result.  While this is an unphysical model, it serves to
point out that while pure reddening is unlikely to reach this part of
the colour-colour diagramme, an old stellar population does 
that.  We thus conclude that EROs which are selected by $r-K>5.5$
(i.e.\ $R-K>5$) which are not EROs by $i-K>4.4$ (i.e.\ $I-K>4$) are
mostly early type galaxies.  In fact, at $i-K<4.2$ there should not be
any contamination from dusty galaxies (compare the vertical dotted
line to the dusty model curves).   

We also plot a sample of averaged SEDs of our EROs in
Fig.~\ref{eroseds}.  The ``excess'' EROs with $i-K<4.2$ are plotted
with squares and are well fit by very evolved galaxy SEDs, as just
discussed.  The critical observed band is the $i$ band, where the
difference between elliptical and dusty starbursts (M82 plotted) is
the largest at the redshifts of our 
targets. Note that the $g$-band is biased to the bluest sources,
since the average is calculated from those EROs detected in the
corresponding band, and only 25\% of the sources have a $g$-band
detection. 

The $r-K$ and $i-K$ selections alone are quite different (see e.g.\
Scodeggio \& Silva \cite{Sc00}).  At least with bright ERO surveys
such as ours, it seems that $r-K$ selects more passive systems. This 
is consistent with spectroscopic followup of $R-K>5.3$ surveys finding
significant amounts of ellipticals (e.g.\ Cimatti et al.\ \cite{Ci02})
and those following up $I-K>4$ selections finding much more star
formation (e.g.\ Yan et al.\ \cite{Ya04}).

\subsubsection{Number densities}

There are 33 ``excess'' $r-K$ EROs with $i-K<4.4$.  They appear to be
at redshifts $z=0.7-1.2$: BC SED fitting favours redshifts below unity
(Fig.~\ref{zphot}), while the GRASIL models suggest redshifts
at $z\approx1-1.2$ (eg.\ Fig.~\ref{ikrk}). These EROs have typical
magnitudes of $K\approx17.2$, 
which would make them at least $\sim2L^{\star}$ galaxies, taking into
account passive evolution since that redshift.   
Very simply, assuming that this group of
EROs consists of ellipticals and is constrained in the mentioned
redshift range 
we derive a co-moving volume of $2\times10^{6}$ Mpc$^{3}$ for the
Mt.Hopkins survey area, and thus a space density of
$\sim2\times10^{-5}$ Mpc$^{-3}$.  
Taking $K$-band LF parameters from Kochanek et al.\ (\cite{Ko01}),
this is a significant part of the expected density of
$\approx8\times10^{-5}$ Mpc$^{-3}$ for local $>2L^{\star}$ early type
galaxies.  Given the incompleteness in our survey at $K>17$, and that
we did not include ellipticals at $i-K>4.4$ in this calculation 
these ERO counts are a very conservative lower limit,
and thus indicate that a major part of massive present day $\sim
2-3L^{\star}$ ellipticals were in place at $z\sim1$.  

The space density just derived is intriguingly close
to those estimated for the higher redshift sub-mm population, which
are suspected to be the most luminous starbursts, perhaps the
progenitors of present day massive ellipticals (see e.g.\ discussion
in Scott et al.\ \cite{Sc02}).

\subsection{Constraints on galaxy formation scenarios}

Figure~\ref{allcounts} plots our ERO counts along with several other
recent surveys and pure luminosity evolution (PLE) models of Daddi et
al. (\cite{Da00}).  All curves depict ellipticals with a $\tau =
0.1$ Gyr initial starburst, and passive evolution afterwards.  The
different lines have formation redshifts of $z_{\rm form} = 2.4, 3.0,$
and 10.  All the models use a Marzke et al.\ (\cite{Ma94}) LF, except
the uppermost dashed line, which instead uses a 2MASS based LF
(Kochanek et al. \cite{Ko01}) resulting in a higher normalization by
approximately a factor of two.

We stress that {\em all EROs are counted} in the figure, i.e.\ no
corrections for the dusty population have been made. As was seen
previously, this is perfectly reasonable for our very bright sample of
EROs.  However, even a very large correction (not supported by our
observations) of 50 \% of the EROs being in fact dusty, would not
change the conclusion significantly.  At somewhat fainter magnitudes,
recent studies have shown (Cimatti et al. \cite{Ci02}, Smail et al.\
\cite{Sm02}) that the percentage of dusty EROs is likely somewhere
around 30--60\%. 

The PLE model using a formation redshift of $z_{\rm form}=3$ fits all
the data 
remarkably well.  Formation redshifts slightly lower than this also
fit well our bright ERO counts, but start underpredicting the numbers
at fainter 
magnitudes. On the other hand, corrections for dusty galaxies might be
larger at the fainter levels (see eg.\ Smith et al.\ \cite{Smi02}).
However, models with $z_{\rm form}<2.4$ start underpredicting the
counts at all levels, especially when using the $R-K>5.3$ selection
criteria.  The very highest formation redshifts of $z_{\rm form}=10$
and over, predict steeper number counts and are more difficult to fit
to both the bright and faint end of ERO counts with the same
normalization.  The figure shows two different LFs giving a factor of
$\sim2$ difference.  Note also that eventhough the $R-K>5.3$
criterion eliminates nearly all the (possibly) lower redshift
$z\sim0.8$ EROs, there is no significant difference in the fits of PLE
models in the right and left panels of Fig.~\ref{allcounts}. 

Using the 2MASS normalized LF it is in fact slightly more
difficult to fit both $R-K>5$ and $R-K>5.3$ counts
simultaneously.  The $R-K>5$ counts are overpredicted by a factor of
$1.5-3$ with all except the very lowest $z_{\rm form}=2.2-2.4$
formation redshifts.  Good fits in the range $K=16.5-18.5$ are again
acquired with $z_{\rm form}\ge4$, but the faintest counts are
overpredicted.  Roche et al.\ (\cite{Ro02},\cite{Ro03}) find their
2MASS-LF based PLE models significantly overpredicting $I-K>4$ ERO
counts, and show that certain amount of merging and density evolution
fits the counts much better in particular in the fainter $K>20$
regime.  

That a range of formation redshifts are necessary to model EROs
is seen for example by comparing Figs.~\ref{ikrk} and~\ref{allcounts}
(see also Cimatti et al.\ \cite{Ci03}): though models with formation
redshifts of $z\approx3$ fit well the {\em counts}, the {\em colours}
produced by such a scenario are not red enough for many EROs found in
this survey, and others.  Very red colours of $r-K>6.5$ or $i-K>5$ can
not be produced without pushing $z_{\rm form}$ closer to 10 (unless
all such extreme EROs are heavily reddened starbursts).

We do not investigate hierarchical formation 
scenarios in more detail here. We merely point out that predictions
from some recent models (Cole et al.\ \cite{Co00} as presented in
Smith et al.\ \cite{Smi02}), fall short an order of
magnitude in the numbers of EROs in the range
$K=17-20$.  See also e.g.\ discussion by Martini (\cite{Ma01}) and
Firth et al.\ (\cite{Fi02}).
It is important to realize that a drastic increase in the
fraction of dusty EROs does not make the hierarchical models fit the
ERO counts any better: they predict too few EROs of all kinds.
On the other hand large fractions of dusty EROs would have to be
subtracted from PLE models including only ellipticals, making all the
Daddi et al.\ PLE models used above to 
overpredict the counts by factors of $1.5-3$.  

It is concluded that the PLE models do give remarkably good fits to
the brightest ERO counts.  Formation redshifts around $z\sim3$ are
favoured -- however, by altering the details of the models, formation
redshifts between $z=2-10$ are also consistent.  Moreover, a range in
the formation era of ellipticals is suggested by the range in ERO
colours.

\section{Summary}

We have searched for EROs in a near-IR survey performed in ELAIS
fields, using $r-K>5.5$, $r-K>5.8$, and $i-K>4.4$ colour criteria.
These are equivalent to the commonly used $R-K>5$, $R-K>5.3$, and
$I-K>4$.  In the survey, reaching approximately $K=17.5$, we find
54 EROs.  The area covered is 2850 arcmin$^2$.

Taking advantage of overlapping mid-IR data, we search for dusty EROs,
since only these should be detected with the used $15\mu$m ISOCAM
band. Only one is found from our conservatively constructed catalogue.
Taking into account 
detection limits we limit the number of very strong starbursts
($SFR\geq200 \, {\rm M_{\sun} year^{-1}}$) in the bright $K<17-17.5$
ERO population to $<10$\%.  

We also make use of a $J-K$ vs.\ optical-infrared colour-colour
diagramme to separate EROs, and find that the fraction of dusty ERO
population is $<10-40$\%, depending on the colour used.  There are
more dusty galaxies in the $i-K$ based ERO selection than if $r-K$ is
used.  HYPERZ photometric redshifts and template fits are also
  employed: 
nearly all redshifts are in the range $z=0.6-1.8$ with a strong peak
at $z\sim0.8$.  Approximately 90\% of the best-fit SEDs are those
of evolved stellar populations. 

We find a considerable amount, $\sim60$\% of all our EROs, of
$r-K>5.5$ EROs which are not EROs with the $i-K>4.4$ criterion.  Using
models we interpret these to be early type galaxies at redshift of
$z\sim0.7-1.1$.  They are interpreted to be the counterparts of local
$2-3L^{\star}$ galaxies, and their resulting space density is
approximately $2\times10^{-5} \, {\rm Mpc}^{-3}$.

Cumulative number counts are provided for the EROs, extending the
available ERO counts to brighter limits than previously.  Our 
counts are consistent with literature counts in the overlapping
magnitude region with same colour cut-offs.  

Our ERO number counts, as well as other literature data, are well
fit by pure luminosity evolution models.  Formation redshifts for
early type galaxies in excess of $z=2.5$ are required to fit the ERO
counts, and  $z\approx3$ is favoured.  However, the range in
the colours of EROs suggests also a wide range in formation redshifts.

\begin{acknowledgements}

We thank the referee for good and clarifying comments. We are grateful
for useful discussion with Kalevi Mattila, Emanuele 
Daddi, and Margrethe Wold.  Emanuele Daddi is especially thanked for
providing the ERO count models used in this paper.

\end{acknowledgements}

\end{document}